\documentclass[journal=jacsat,manuscript=article]{achemso}

\usepackage[version=3]{mhchem} 
\usepackage{siunitx}
\usepackage[markup=underlined]{changes}
\usepackage{verbatim}
\colorlet{Changes@Color}{red}
\usepackage[frak=esstix]{mathalpha}

\usepackage{graphicx} 
\usepackage{amsmath}

\title{Infrared Metaplasmonics} 
\author{Zarko Sakotic}
\email{zarko.sakotic@austin.utexas.edu}
\author{Noah Mansfeld}
\author{Amogh Raju}
\author{Alexander Ware}
\author{Divya Hungund}
\author{Daniel Krueger}
\author{Daniel Wasserman}
\affiliation[UT Austin]{The Chandra Department of Electrical and Computer Engineering, University of Texas at Austin, Austin, TX 78758, United States}
\email{dw@utexas.edu}
\date{December 11th 2024}

\begin{document}

\maketitle

\begin{abstract}
   Plasmonic response in metals, defined as the ability to support subwavelength confinement of surface plasmon modes, is typically limited to a narrow frequency range below the metals' plasma frequency.  This places severe limitations on the operational wavelengths of plasmonic materials and devices. However, when the volume of a metal film is massively decreased, highly confined quasi-two-dimensional surface plasmon modes can be supported out to wavelengths well beyond the plasma wavelength.  While this has, thus far, been achieved using ultra-thin (nm-scale) metals, such films are quite difficult to realize, and suffer from even higher losses than bulk plasmonic films. To extend the plasmonic response to the infrared, here we introduce the concept of metaplasmonics, representing a novel plasmonic modality with a host of appealing properties. 
   By fabricating and characterizing a series of metaplasmonic nanoribbons, we demonstrate large confinement, high quality factors, and large near-field enhancements across a broad wavelength range, extending well beyond the limited bandwidth of traditional plasmonic materials. We demonstrate $35\times$ plasmon wavelength reduction, and our numerical simulations suggest that further wavelength reduction, up to a factor of 150, is achievable using our approach. The demonstration of the metaplasmonics paradigm offers a promising path to fill the near- and mid-infrared technological gap for high quality plasmonic materials, and provides a new material system to study the effects of extreme plasmonic confinement for applications in nonlinear and quantum plasmonics.
\end{abstract}

\section{Introduction}


Harnessing the interaction between light and free electrons in materials is the basis of plasmonics, a transformative field in modern optics and materials science\cite{maier2007plasmonics}. The singularly important and defining aspect of plasmonics is the ability to confine light at the nanoscale, far below the scale of the photon wavelength itself, offering a path towards overcoming the mismatch between photon and electron length scales. While noble metals (e.g. \ce{Au}, \ce{Ag}) are the exemplary materials enabling such confinement, true plasmonic response is only possible at frequencies just below these metals' plasma frequency, where the metal permittivity is negative, but of the same order as the surrounding dielectric material. At longer wavelengths (lower frequencies) the noble metals behave more like perfect electrical conductors (PECs), and subwavelength plasmonic confinement is not possible\cite{SLawNP}. Efforts to extend plasmonic response to longer wavelengths 
have generally focused on alternative plasmonic materials with reduced carrier concentrations, such as transparent conducting oxides or doped semiconductors\cite{naik2013alternative, law2012mid,taliercio2019semiconductor}. However these materials' figures of merit do not exceed those of noble metals\cite{hsieh2017comparative,pellegrini2018benchmarking,zhong2015review}, and optical confinement still scales with wavelength, precluding nanoscale mode confinement. Recently, it has been demonstrated that plasmonic response in conducting materials can be dramatically extended when one of the dimensions of the plasmonic material is reduced to the atomic scale \cite{basov2016polaritons,rivera2016shrinking,li2017plasmonics}. Due to this effect, the rise of 2D materials such as graphene\cite{geim2007rise} has substantially expanded the impact of plasmonics, in particular for opto-electronic and nanophotonic applications, due to the dramatically enhanced light-matter interaction and extraordinary tunability achievable in these 2D material systems \cite{bonaccorso2010graphene,grigorenko2012graphene,garcia2014graphene}. However, the low carrier concentration of graphene limits its appealing plasmonic properties to the THz and longer wavelength regions of the mid-infrared (mid-IR, 3-\SI{14}{\um}) \cite{yan2013damping,brar2013highly}, and broad-area scaling of graphene-based optoelectronics remains a challenge. This leaves a rather significant technological gap, extending from the near-infrared (NIR, 0.75-\SI{1.4}{\um} through to the long wave infrared (LWIR, $8-14 \mu m$), for plasmonic materials with large figures of merit i.e. strong confinement and high quality factors.

Following in the footsteps of graphene, the recent emergence of atomically thin metals has expanded the plasmonic palette and extended plasmonics into the technologically-vital short-wave infrared (SWIR, 1.4-\SI{3}{\um}) \cite{de2015plasmonics,rivera2016shrinking,maniyara2019tunable,abd2019plasmonics,yakubovsky2019ultrathin,yakubovsky2023optical,karaman2024ultrafast,pan2024large,mkhitaryan2024ultraconfined,novoselov2024graphene}, offering one promising path to fill this gap. The extreme confinement in these few-nm thick films has enabled electrical tunability\cite{manjavacas2014tunable,maniyara2019tunable}, strongly enhanced nonlinear response\cite{pan2024large,lv2024voltage} and Purcell effect\cite{mkhitaryan2024ultraconfined}, along with observable quantum effects\cite{qian2016giant}. Ultimately, however, the applicability and further improvement of 2D metal plasmonics depends on the scalability, reproducibility, and optical/electrical quality of the metal films themselves, as is the case with graphene\cite{boggild2023research}. While the reported advances in 2D metal film synthesis are substantial, the complex fabrication and patterning techniques remain an impediment to their wide applicability. Perhaps more fundamentally, the atomically thin nature of these metal films precludes them from maintaining bulk optical quality, due to the intrinsic electron confinement effect, i.e. Landau damping, that emerges at these length scales\cite{fuchs1938conductivity,kreibig2013optical,khurgin2015ultimate,khurgin2017landau} even for perfect, atomically smooth films\cite{maniyara2019tunable,martinez2021ultrathin,akolzina2024optical}. Thus, these intrinsic limitations seemingly hinder the wavelength extension, scalability, and thus, potential impact, of ultra-thin metals in infrared plasmonics. 

Here we propose and demonstrate the concept of metaplasmonics to address this issue and achieve high-quality quasi-2D metal plasmonics across the SWIR and mid-IR wavelength ranges. By leveraging an ultra-thin (but not so thin as to incur Landau damping effects), highly-diluted metal film, we demonstrate quasi-two-dimensional plasmons with bulk optical quality from the NIR to the LWIR, a dramatic extension of plasmonic response with implications for a wide range of infrared applications. By lithographically perforating a thin metal on an ultra-subwavelength scale, but in such a way that no features are smaller than the electron mean free path, we can massively decrease the active plasmonic volume (i.e. the effective permittivity of the metal) without sacrificing optical quality, unfolding the large wavevector space otherwise only attainable with atomically thin metal layers. The ultrathin metaplasmonic film is characterized with salient features including broadband operation, low loss, extreme confinement, and large electric field enhancement. The near- and far-field properties of the metaplasmonic films exceed those of the nanometric mono- and poly-crystalline metal films due to substantially lower scattering rates, while they can be produced/scaled with standard deposition and lithography methods. We also show that our approach can be extended to arbitrarily long wavelengths, offering an alternative to graphene or doped-semiconductor plasmonics.

\section{Results}

\begin{figure} [t]
\centering
\includegraphics[width=14cm]{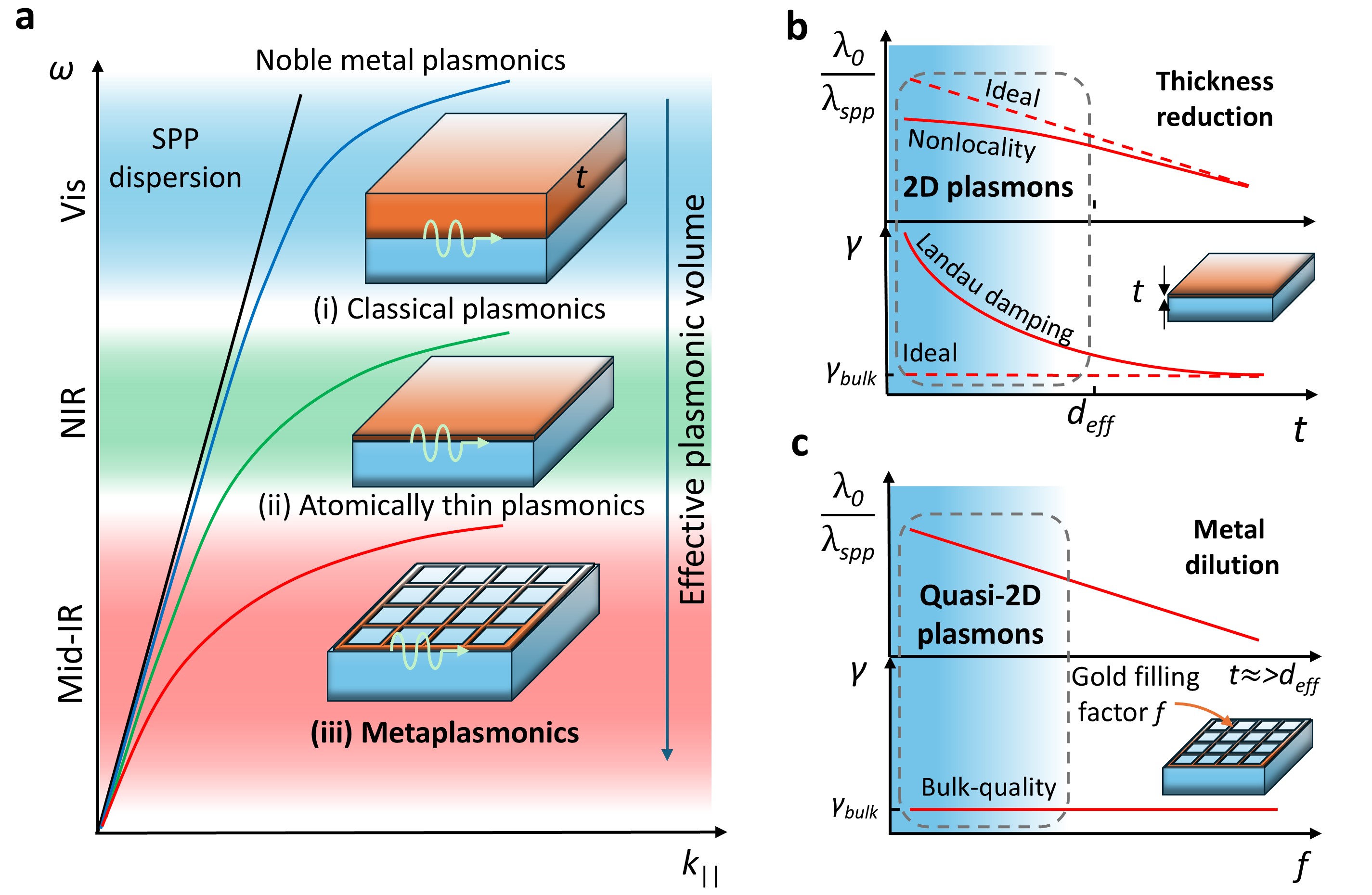}
\caption{(a) Conceptual schematic and SPP dispersion of (i) classical, (ii) ultrathin, and (iii) metaplasmonic approaches. As the volume of the metal film is decreased, the bending of the dispersion, characteristic of plasmonic response, moves to lower frequencies.  Fabrication limitations and Landau damping limit this shift for thin metal films, however the proposed metaplasmonic approach can extend plasmonics well into the mid-IR. (b) In atomically-thin films the confinement factor increases linearly with decreasing thickness. However, at thicknesses well below $d_{eff}$ Landau damping dominates the optical response, strongly increasing scattering loss. (c) In the proposed metaplasmonic film, reducing the metal fill fraction enables an analogous confinement effect as in (b), but eliminates the effects of Landau damping by keeping all the dimensions on the scale of $d_{eff}$ or larger, preserving bulk scattering rates.}
\label{Figure:1}
\end{figure}

The proposed metaplasmonic concept is depicted in Fig. \ref{Figure:1}a, which shows the dispersion for the classical plasmonic interface between a thick (or semi-infinite) noble metal and a dielectric superstrate. For such a system, the surface plasmon polariton (SPP) diverges from the light line only at high frequencies, near the noble metal plasma frequency (Fig. \ref{Figure:1}a-i). The same schematic shows the shift of the plasmonic dispersion achievable by replacing the thick metal film with an ultrathin metal deposited upon a dielectric substrate (Fig. \ref{Figure:1}a-ii).  
Using the local response approximation, and when the metal film is sufficiently thin, the plasmon wavevector for system can be approximated as\cite{abd2019plasmonics}:

\begin{equation}{k_{spp}(\omega)} \approx \frac{(\epsilon_s+1)}{(1-Re[\epsilon_m(\omega)])t}.\label{eqn:1}\end{equation}

where $\epsilon_m$ and $\epsilon_s$ are the metal and substrate permittivity, and $t$ is metal thickness. In the case of classical plasmonics, when the film is a few tens of nanometers thick, the strongly confined i.e. large $k_{spp}$ modes emerge only when the metal permittivity is small (for noble metals, this occurs at visible wavelengths). When the metal thickness is reduced to few nanometers\cite{maniyara2019tunable}, the dispersion diverges from the light line, according to Eq. \ref{eqn:1}, at much lower frequencies (Fig. \ref{Figure:1}a-ii), enabling strong mode confinement into the SWIR. This simple but remarkable property was only recently demonstrated, enabled by the ability to produce continuous, atomically-smooth films of a few nm in thickness. However, when inspecting Eq. \ref{eqn:1}, an alternative mechanism for increasing the SPP wavevector becomes apparent: reduction of the metal permittivity. Such a reduction can be achieved via metal dilution, by which we can produce a completely analogous bending of the plasmon dispersion;  such a metaplasmonic film, with plasmonic response out to the mid-IR, is depicted in Fig. \ref{Figure:1}a-iii. We have recently demonstrated that ultrathin perforated metal films can fully mimic optical properties of planar, atomically thin films\cite{sakotic2024mid} which gives initial credence to this idea, although no surface modes were considered in that work.

While at first glance thickness reduction and metal dilution seem to be equally acceptable approaches to minimizing plasmonic volume (at least in this simplified picture), the key difference between these approaches lies in the imaginary part of the metal permittivity i.e. the scattering rate of free carriers in the metal. As illustrated in Fig. \ref{Figure:1}b, when metal thickness decreases, the confinement factor $\lambda_{0}/\lambda_{spp}$ linearly increases \cite{abd2019plasmonics}, reaching the 2D plasmon range at a few nm in thickness. However, as the thickness drops below the electron mean free path length ($d_{eff}\simeq10-\SI{20}{\nm}$ for bulk \ce{Au}\cite{olmon2012optical}), the electrons scatter at the metal/dielectric interfaces more frequently, causing Landau damping, which quickly becomes the dominant damping mechanism at these length scales\cite{fuchs1938conductivity,kreibig2013optical,khurgin2015ultimate,mortensen2021mesoscopic}. In this regime, at thicknesses of a few nanometers, the scattering rate roughly follows $\gamma\approx v_F/t$ \cite{kreibig2013optical}, where $v_F=1.4 \times 10^6\SI[per-mode=symbol]{}{\meter\per\second}$ is the Fermi velocity of electrons in bulk gold. This is the case even in atomically-smooth, fully continuous films, where other loss mechanisms such as grain-boundary and surface-roughness scattering are minimized\cite{martinez2021ultrathin}.  Landau damping causes the scattering rate to increase (from the bulk) by an order of magnitude when film thicknesses of a few nanometers are reached\cite{akolzina2024optical}. This severely lowers such films' quality factors $Q=\omega/\gamma$, and imposes a strict trade-off between confinement and Q, particularly deleterious for resonances in the mid-IR\cite{maniyara2019tunable,abd2019plasmonics}. While nonlocal effects can also be a limiting factor due to the surface plasmon polariton (SPP) confinement saturation\cite{khurgin2015ultimate}, in the case of atomically-thin films Landau damping by far represents a greater impediment to low loss and high confinement plasmonics than nonlocality, at least in the regimes studied thus far.

Alternatively, by starting with a metal film whose thickness is on the scale of, or slightly thicker than, the electron mean free path $d_{eff}$, and subsequently perforating the film so as to minimize the metal filling factor without introducing features smaller than $d_{eff}$ (in either lateral or vertical dimensions), we can circumvent the effects of Landau damping and preserve the bulk optical quality of the metal, as shown schematically in Fig. \ref{Figure:1}c. Thus we can reduce the metal filling factor $f$ and achieve strong confinement without ever increasing the scattering rate $\gamma$. This is the key to unlocking both large $k_{spp}$ and high Q-factors across NIR, SWIR, and mid-IR wavelengths.

\begin{figure} [t]
\centering
\includegraphics[width=16cm]{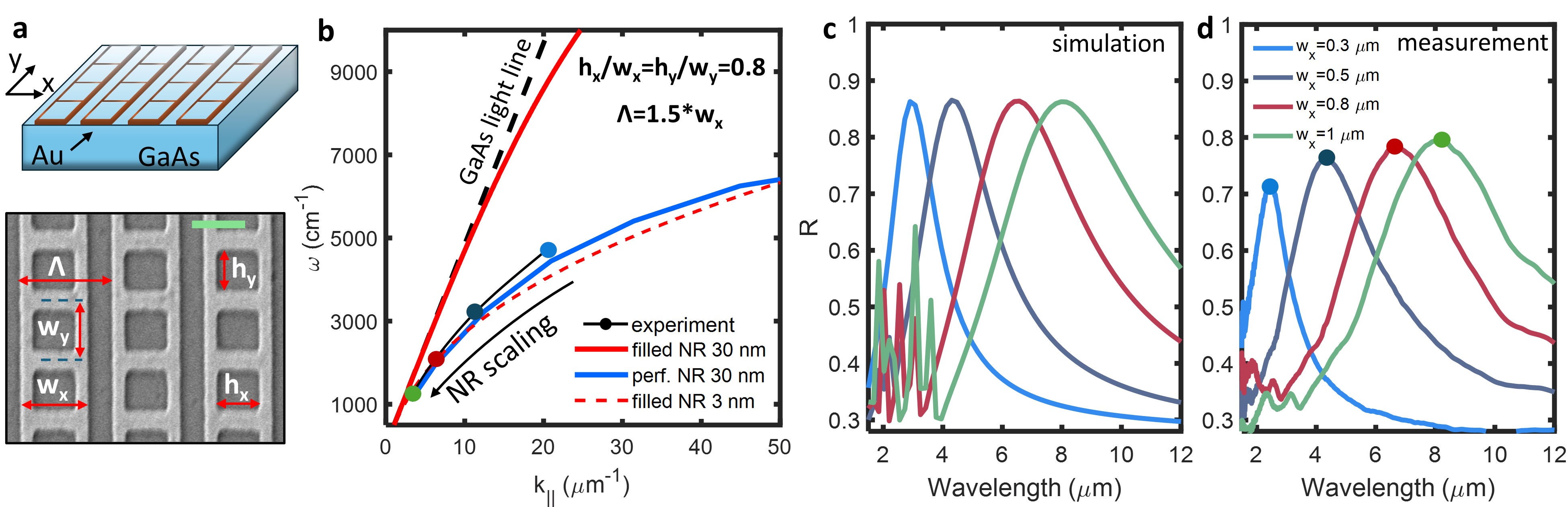}
\caption{(a) Schematic and plan view scanning electron micrograph of the fabricated \SI{30}{\nm} thick metaplasmonic nanoribbon with NR period ($\Lambda$), width ($w_x$), and hole width and height $h_x$, $h_y$; the scale bar is $\SI{400}{\nm}$. (b) Simulated SPP dispersion of the filled ($h_x=0$) and perforated NRs, where the \SI{30}{\nm} filled NR shows weak confinement, while the perforated NR and \SI{3}{\nm} filled NR show strong plasmonic dispersion. To obtain the dispersion the lateral size of the NRs $w_x$ is scaled with constant period to width ratio $\Lambda/w_x=1.5$, with square perforations such that $h_x/w_x=0.8$ is constant. (c) Simulated and (d) experimental infrared reflection spectra as the lateral size ($w_x$) of the metaplasmonic NR is scaled.}
\label{Figure:2}
\end{figure}

To harness these features and enable access to the highly confined modes of the metaplasmonic film from free space, some sort of structuring is necessary. Thus, for an initial experiment, we deposit metaplasmonic nanoribbons (NR), using electron-beam lithography (EBL), metalization, and lift-off, on semi-insulating GaAs substrates, as shown schematically in Fig. \ref{Figure:2}a. A scanning electron microscope (SEM) image of one of the fabricated NRs is shown in Fig. \ref{Figure:2}a (note that all dimensions considered are larger than $d_{eff}$ and significantly less than the free space wavelengths of interest). The structuring of the metaplasmonic film enables coupling into SPP modes localized on the NRs, for TM-polarized light (electric field across the NRs). To better understand the nature of the SPP modes, we first numerically compute the dispersion of the metaplasmonic NRs with thickness $t=\SI{30}{\nm}$ in Fig. \ref{Figure:2}b. In the quasi-static approximation, scaling the size of NRs while keeping the period-to-width ratio constant traces the dispersion of the extended film's modes where $k_{spp}\approx\pi/w_x$ and $\lambda_{spp}\approx2 w_x$\cite{yan2013damping,maniyara2019tunable}. In the case of a filled NR (no perforation, red line), the standard antenna-like dipolar mode with weak confinement tightly follows the light line throughout the NIR, SWIR, and mid-IR regions. When a square hole of the size $h_x/w_x=h_y/w_y=0.8$ is used to perforate the NR, the mode significantly bends away from the light line (blue line), indicating much stronger confinement across the SWIR and mid-IR.    

The excitation of these localized plasmonic modes is measured by infrared reflection spectroscopy under normal incidence, as shown in Fig. \ref{Figure:2}c,d. To probe the ability of the metaplasmonic NR to support plasmonic response across a broad range of wavelengths, we scale the size of the NRs from $w_x=\SI{300}{\nm}$ to $w_x=\SI{1000}{\nm}$, with $h_x/w_x=0.8$ and $\Lambda/w_x=1.5$, which traces the dispersion from Fig. \ref{Figure:2}b. The experimental and simulated results show excellent agreement, although there is a small blue shift of the measured data, a result of the discrepancy between the EBL design and the final, fabricated metaplasmonic geometry (SI). The NR plasmon, although moderately confined in this example, well surpasses the conventional wavelength-scale noble metal plasmonic modes in the mid-IR. The reflection peaks are large, indicating very small absorption and low loss, while the plasmon resonance widths show an excellent match to the simulation, confirming that the film preserves the bulk optical quality of the deposited gold. Additionally, we also plot the dispersion of a $t=\SI{3}{\nm}$ thick gold nanoribbon (red dashed line, Fig. \ref{Figure:2}b), which shows analogous dispersion bending and confirms our prediction that diluted films mimic the plasmonic behavior of much thinner films; however the reflection peaks drop significantly in amplitude for the $\SI{3}{\nm}$ continuous metal NR, and are much broader when the electron confinement effect is accounted for. 

\begin{figure} [t]
\centering
\includegraphics[width=16cm]{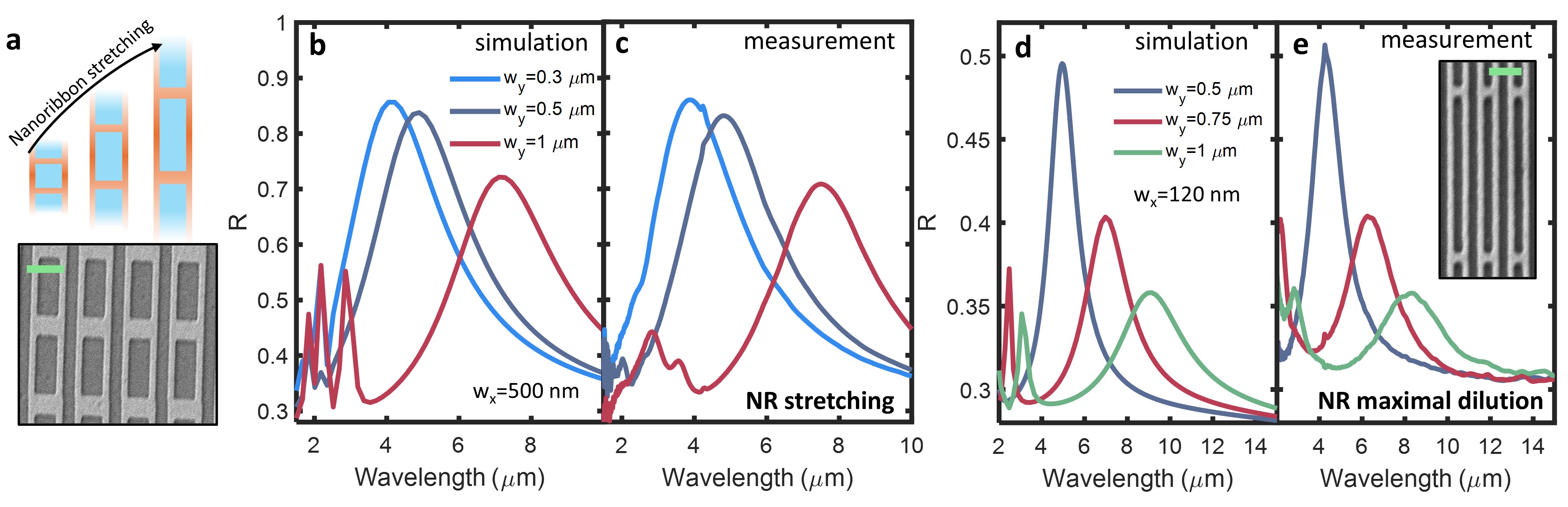}
\caption{(a) Schematic of nanoribbon stretching and an SEM image of a stretched NR array with scale bar $\SI{500}{\nm}$. (b) Simulated and (c) measured infrared reflection spectra as the y-period of the NRs ($w_y$) is scaled, while other geometrical parameters are fixed at $w_x=\SI{0.5}{\um}$, $\Lambda/w_x=1.5$ and $h_x/w_x=h_y/w_y=0.8$. (d) Simulated and (e) measured infrared reflection spectra for maximal dilution, achieved by increasing y-period ($w_y$) with constant lateral size of the gold features: $w_x - h_x = \SI{40}{\nm}$ and $w_y - h_y = \SI{30}{\nm}$ in x- and y-directions; NR width fixed at $w_x= \SI{120}{\nm}$ with thickness $t= \SI{25}{\nm}$. Inset: SEM image of the maximally diluted and stretched metaplasmonic NR; scale bar is $\SI{200}{\nm}$.}
\label{Figure:3}
\end{figure}

Due to the structured nature of our metaplasmonic film, the design space allows us to choose the periodicity of the perforation in the direction parallel to the NRs (in this case along the y-axis), as sketched in Fig. \ref{Figure:3}a, which can tune the spectral position of the plasmon resonance. The `stretching' of the NR increases the length and thereby the inductance of the edge lines which dilutes the film's effective permittivity further\cite{pendry1998low}, moving the plasmon resonance to longer wavelengths i.e. increasing the confinement factor. 
To test this appealing property in experiment, we fabricate a series of NRs where we keep the NR unchanged in the x-direction (i.e. the $k_{spp}\approx\pi/w_x=2\pi\SI{}{\um}^{-1}$ is kept constant, with constant filling factors $h_x/w_x=0.8$ and period-to-width ratios $\Lambda/w_x=1.5$), while we change the y-periodicity from $w_y=\SI{300}{\nm}$ to $w_y=\SI{1000}{\nm}$, Fig. \ref{Figure:3}b,c. Our experimental results closely match the simulations, showing that we are able to tune the NR plasmon from $\SI{3.8}{\um}$ to $\SI{8}{\um}$, leading to much stronger confinement of the plasmon $\lambda_{0}/\lambda_{spp}=8$. 

In order to test even stronger confinement regimes, we have also fabricated $w_x=\SI{120}{\nm}$ wide NRs ($k_{spp}\approx\SI{26.2}{\um}^{-1}$), but contrary to scaling of the whole unit cell in y-direction as in the previous example, we set constant, minimal gold feature sizes of \SI{40}{\nm} and \SI{30}{\nm} in the x- and y-directions, respectively. In this way, as $w_y$ is increased, the metal volume is minimized, which should maximize confinement. Indeed, the measured spectra indicates that the spectral position of the plasmonic resonance increases substantially, from $\SI{4.2}{\um}$ ($w_y=\SI{500}{\nm}$) to $\SI{8.5}{\um}$ ($w_y=\SI{1000}{\nm}$), demonstrating a remarkable confinement factor, going from $\lambda_{0}/\lambda_{spp}=18$ to $\lambda_{0}/\lambda_{spp}=35$. This is a stark difference compared to the confinement factor of conventional noble metals in the infrared $\lambda_{0}\approx\lambda_{spp}$, and is even larger than the maximal confinement factor of 15 for \ce{Ag} (which is considered the best overall plasmonic material) in the visible range\cite{jablan2009plasmonics}.
Moreover, the metaplasmonic approach points to the potential of reaching the extreme confinement regime $\lambda_{0}/\lambda_{spp}>100$, something that has only been demonstrated with true 2D materials such as graphene or van der Waals layers thus far; this is discussed further later in the manuscript as we reflect on the limits of our approach. Nevertheless, these results indicate that the metaplasmonic film represents a broadband infrared plasmonic material with excellent confinement and material quality factor. 

\begin{figure} [t]
\centering
\includegraphics[width=16cm]{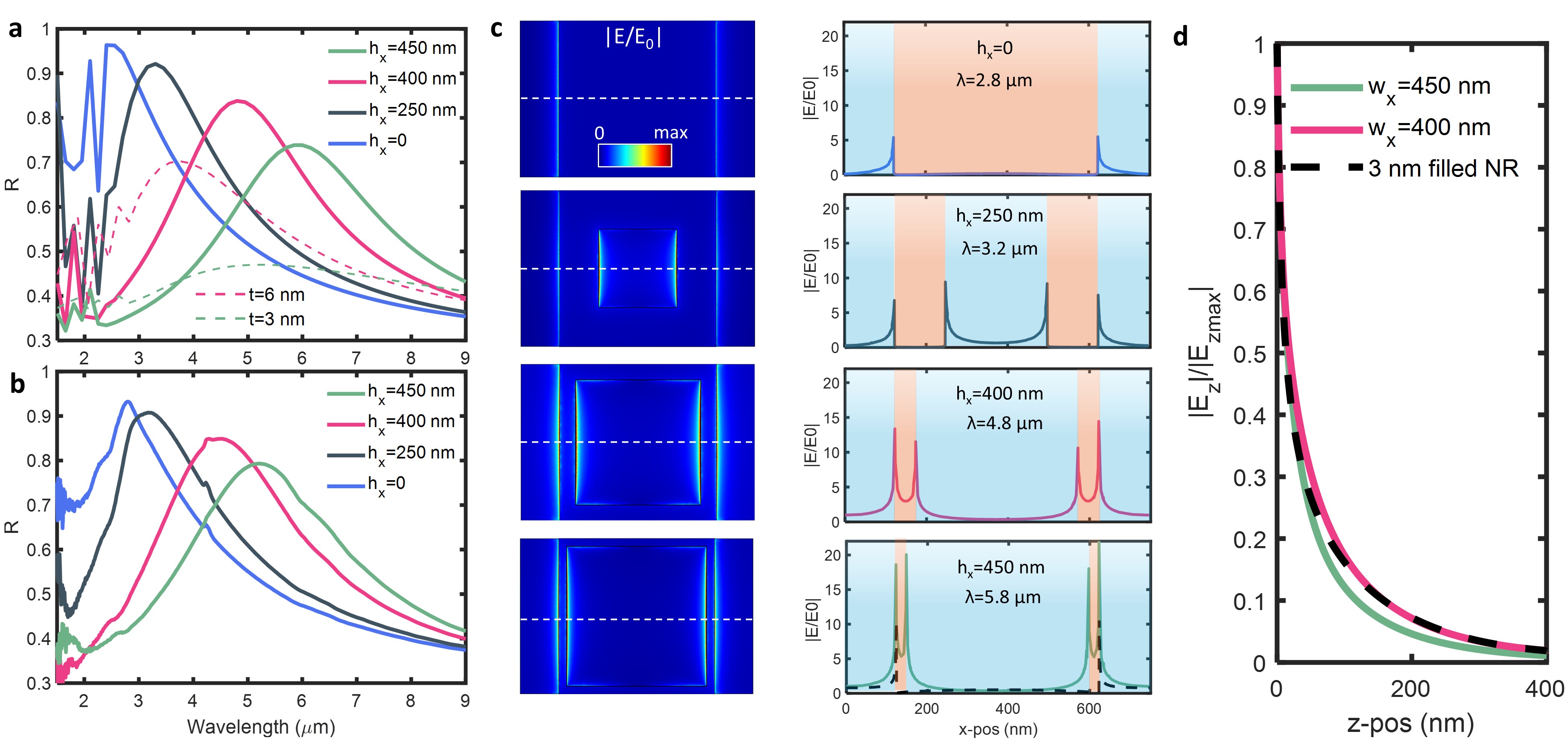}
\caption{(a,b) Simulated and experimental spectra for perforated NRs with increasing dilution. Simulated response of filled NRs (dashed lines) with $t=\SI{6}{\nm}$ and $t=\SI{3}{\nm}$ for comparison. (c) Left column is the plan-view of the electric field for the four dilution cases; right column represents the electric field enhancement plotted in the middle of the unit cell along the x-axis;  (d) Simulated electric field decay (z-component, normalized to the maximum) along the z-direction for $w_x= \SI{450}{\nm}$, $w_x= \SI{400}{\nm}$, and the filled $t= \SI{3}{\nm}$ NR.}
\label{Figure:4}
\end{figure}

To fully appreciate the effects of dilution and better understand both the far-field and near-field nature of the metaplasmonic NR plasmons, we fabricated and analyzed $t=\SI{30}{\nm}$ thick gold NRs of width $w_x=\SI{500}{\nm}$ and period $\Lambda=\SI{750}{\nm}$, with varying widths of the patterned perforation ($h_x$). In this manner we vary the extent to which the NR metal is diluted. Fig. \ref{Figure:3}a, and b show the simulated and experimental reflection of the metaplasmonic NRs with varying dilution, while Fig. \ref{Figure:4}c shows the plan-view electric field profile of the NRs on resonance. Considering first the film with no dilution ($h_x=\SI{0}{\nm}$), we observe the first dipolar mode of the nanoribbon at $\lambda_o = \SI{2.8}{\um}$. The field profile, on resonance, of the undiluted metal ribbon displays only the modest confinement and field enhancement expected of what is essentially a PEC dipole antenna. By increasing the hole size, the plasmon resonance $\lambda_{0}$ moves towards longer wavelengths, indicating increasing confinement factor $\lambda_{0}/\lambda_{spp}$. As discussed earlier, this effect is completely analogous to reducing the thickness of the filled NR; for comparison we also simulate the response of planar NRs with thicknesses $t=\SI{6}{\nm}$ and $t=\SI{3}{\nm}$ (dashed lines, Fig. \ref{Figure:4}a). 
 While analogous plasmon resonance shifting is achieved with reducing thicknesses, when factoring in Landau damping (the thickness-dependent scattering rate reported in the literature) \cite{martinez2021ultrathin,akolzina2024optical,maniyara2019tunable}, the reflection peaks become much weaker and broader compared to the metaplasmonic NRs, indicating lower Q factors and larger loss. Our measurement results closely follow the simulations (which use the bulk optical properties of our metals), demonstrating that bulk quality is preserved even at smallest metal filling fraction, and that quasi-2D NR plasmons substantially outperform atomically thin metal films in terms of confinement and Q-factors. It is also worth noting that for incident waves polarized along the NR (y-polarization), the response of the most diluted NR is only slightly stronger than the substrate background response, indicating that the metaplasmonic film is essentially transparent to infrared waves (SI).

The dramatic wavelength reduction, and strong coupling to free space, of the metaplasmonic structures suggest their suitability as an infrared plasmonic materials system. However, a significant appeal of plasmonics is the ability to strong confine light to subwavelength, nanoscale dimensions. For this reason we also study study the near-field properties of the metaplasmonic NRs. In Fig. \ref{Figure:4}c, we plot the electric field enhancement at the peak wavelength of the plasmon for each dilution case, to capture the near-field enhancement (NFE) present at the edges of the NR. The panels in the left column represent the simulated electric field enhancement plots across the unit cell in the x-y plane, while the right column shows NFE line along the x-axis in the middle of the unit cell. Compared to the unperforated film where the dipolar mode exhibits the usual NFE along the edges of the ribbon, the metaplasmonic NR field is enhanced at both sides of the edge lines, resulting in four peaks along the spatial dimension of the unit cell. For stronger dilution, the NFE starts to rise substantially, resulting in very large peak fields in the most diluted example. The simulated comparison between the most diluted case and the $t=\SI{3}{\nm}$ film shows that the metaplasmonic NR is characterized by significantly larger NFE, as well as providing two additional regions with large field enhancement on the inner edges of the lines. Furthermore, the quasi-2D plasmon in the metaplasmonic NR is very tightly confined in the out-of-plane $z$-direction, where the electric fields for the more diluted NRs decay on a similar lengthscale as the $t=\SI{3}{\nm}$ NR, Fig. \ref{Figure:4}d.

The field distribution reveals an intriguing aspect of the metaplasmonic NR plasmon - as long as the electrical continuity of the NR edges is preserved, the active plasmonic volume of the NR can be reduced to arbitrarily low values leading to plasmons with arbitrarily large levels of confinement. While this reduction in volume will eventually be limited by nonlocal effects, in our metaplasmonic films both the physical dimensions and the level of confinement are far from the onset of Landau damping. This is evident from our simulations based on local response, which faithfully reproduce the experimentally observed reflection. Moreover, while we use nanoribbons as our plasmonic resonator and square/rectangular holes as our dilution mechanism, the metaplasmonic concept can be applied to any localized surface plasmon resonance (LSPR) structure using arbitrarily shaped perforations, as long as the electrical continuity of the resonator edges is preserved.

It is important note here that our proof-of-concept demonstration is far from the limits of our approach; in our experiments the gold thickness is $\simeq\SI{30}{\nm}$ (along with a $\SI{3}{\nm}$ Cr adhesion layer) while the minimal tested lateral feature sizes were also around $\simeq\SI{30}{\nm}$, both of which are several times larger than the mean free path length $d_{eff}$ of bulk gold. It has been shown that producing bulk-like quality films of gold at $t=\SI{9}{\nm}$ is possible\cite{yakubovsky2023optical}, while large-scale $\SI{10}{\nm}$ lithography is readily available using EBL tools\cite{jung2020recent}. Using these as realistic limits of our design space i.e. using minimum feature sizes of $\SI{10}{\nm}$ in all spatial dimensions would allow us to further minimize the plasmonic volume, and thus maximize the confinement factor, while largely avoiding the Landau damping regime. 

\begin{figure} [t]
\centering
\includegraphics[width=16cm]{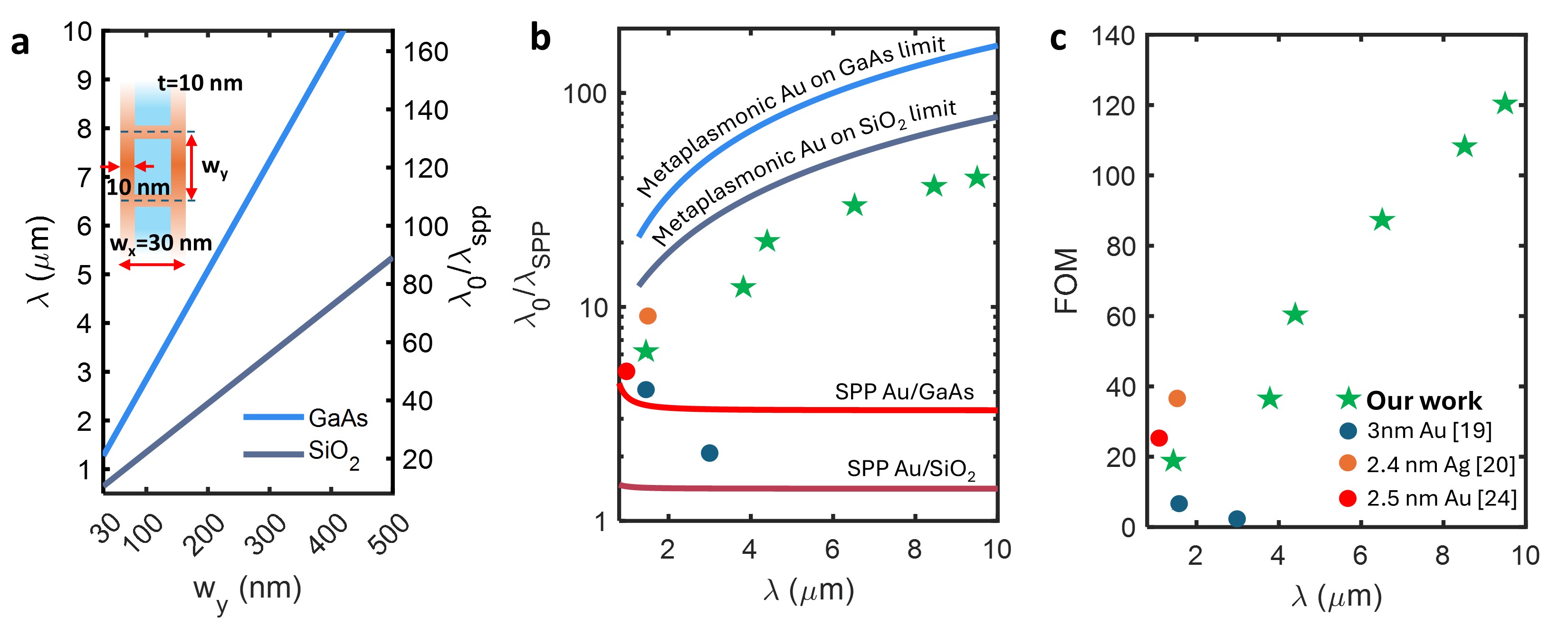}
\caption{(a) Metaplasmonic NR at the limits of the dilution approach for high index (\ce{GaAs}) and low index (\ce{SiO_2}) substrates; Inset shows the NR with \SI{10}{\nm} gold feature sizes in all dimensions. As the y-direction period $w_y$ is increased, the plasmon resonance for metaplasmonic films on \ce{GaAs} is increased from the NIR to the LWIR, covering confinement factors from 20 at $\SI{1.5}{\um}$ to 150 at $\SI{9}{\um}$. For metaplasmonic films on \ce{SiO_2}, the simulated confinement factor goes from 15 at $\SI{0.8}{\um}$ to 80 at $\SI{4.8}{\um}$. 
(b) Confinement factor and (c) FOM ($Q \times \lambda_{0}/\lambda_{spp}$) comparison of demonstrated metaplasmonic NRs and previously realized atomically thin metal plasmonic NRs in the infrared.}
\label{Figure:5}
\end{figure}

 To test this possibility, we employ numerical simulations of the NR operating at the proposed limits of our approach, Fig. \ref{Figure:5}. In this example, we use a $t=\SI{10}{\nm}$ thick NR and fix its width to $\SI{30}{\nm}$ (plasmon wavelength fixed at $\lambda_{spp}\approx\SI{60}{\nm}$), while all the in-plane gold features are fixed at $\SI{10}{\nm}$. The NR plasmon's out-of-plane decay in this example would be slightly smaller than \SI{10}{\nm} ($\SI{60}{\nm}/4\pi\approx \SI{5}{\nm}$), where some Landau damping would appear, although for gold this is not expected be substantial\cite{khurgin2015ultimate}. This allows us to maximally dilute the film while preserving bulk-like quality. Remarkably, our simulations of NRs on \ce{GaAs} operating at these limits indicate that $\lambda_{0}/\lambda_{spp}>150$ can be achieved, Fig. \ref{Figure:5}a (SI); the resonant wavelength increases linearly with the y-period, going from NIR to LWIR as the y-period is increased from $w_y=\SI{20}{\nm}$ to $w_y=\SI{500}{\nm}$. Due to the presumably preserved bulk quality, quality factors remain large, around 6 in the SWIR and over 3 up to the LWIR (SI), only limited by the intrinsic bulk damping. This combination of confinement and quality factors is unprecedented for noble metals anywhere in the infrared regime, and has only been demonstrated by graphene and van der Waals materials thus far\cite{basov2016polaritons,low2017polaritons,jablan2009plasmonics}. Beyond the giant confinement which is achieved in the LWIR, the broadband nature of the metaplasmonic approach allows coverage of wavelength ranges inaccessible to the aforementioned 2D materials i.e. NIR to mid-wave IR ($0.75-5 \mu m$) with strong confinement and large Q-factors. Furthermore, using a low index substrate such as \ce{SiO_2}, we can achieve a factor of two improvement in Q-factor for the same NR, although confinement is weaker by the same factor (SI). Inevitably, however, as is the case with other plasmonic materials, when the confinement factor is increased the plasmon lifetime decreases, which indicates the trade-off between confinement and propagation length associated with any plasmonic material system. Thus, depending on the application and wavelength of operation, one can tailor the amount of dilution for the intended purpose. These results challenge the traditionally held notion that noble metals have poor confinement properties in the infrared due to their prohibitively large permittivity; the metaplasmonics approach provides a recipe for turning a material with large carrier concentration into a massively diluted system with plasmonic properties akin to those of true 2D-materials. 

 To put the discussed limits and the demonstrated results in context, in Fig. \ref{Figure:5}b we compare the confinement factor of the metaplasmonic NRs with NRs based on atomically thin metal films; we also plot the previously discussed limits as well as the SPP confinement factor for classical plasmonic \ce{Au}/\ce{SiO_2} and \ce{Au}/\ce{GaAs} interfaces. In the NIR and SWIR, our metaplasmonic film compares favorably to the 3 nm polycrystalline gold film \cite{maniyara2019tunable}, while it shows similar confinement factor to monocrystalline silver and gold NRs of few nm thickness \cite{abd2019plasmonics,pan2024large}. Metaplasmonic NRs excel in the mid-IR, where no other demonstrations of confined plasmonic modes using ultrathin noble metal NRs have been reported, showing an order of magnitude improvement over the NIR demonstrations and the classical SPP confinement at \ce{Au}/\ce{SiO_2} and \ce{Au}/\ce{GaAs} interfaces. 
 In contrast to atomically thin films, the bulk quality of the metaplasmonic NRs enables large quality factors throughout the infrared. To factor in both the confinement and the quality factors of the resonances, we define a figure of merit (FOM) for the NR performance as  $FOM=Q\times\lambda_{0}/\lambda_{spp}$, shown in Fig. \ref{Figure:5}c. For the maximally dilluted NRs, the measured Q factor remained around $\approx3$ throughout the mid-IR and thus shows very large FOM, which is only possible due to the absence of Landau damping that would otherwise strongly suppress the ultra-confined modes.
 
\section{Discussion}
It is worth noting that the proposed dilution approach reduces the volume of the material while preserving its electrical continuity, which is important so that electron movement and screening of the photon field is not impeded by any discontinuities (both in the extended film or within a finite LSPR resonator). This ensures true plasmonic behavior the metaplasmonic film\cite{pendry1998low}. Furthermore, the films studied here maintain high electrical conductivity, and depending on the the feature sizes and dilution factor the film can also enable optical transparency at wavelengths well below the plasmonic regime (SI), making the metaplasmonic films appealing for transparent electrodes and electrical tunability applications. In that sense, the extended metaplasmonic film in our work resembles the works on transparent nanowire electrodes\cite{van2012transparent}, and more broadly the inductive metal grids \cite{ulrich1967far,luukkonen2008simple}. While plasmonic modes in thin nanowire networks have been studied in the visible and NIR regimes, the associated limits and connection between thickness/dilution and SPP modes were not explored, while no examples of translation into patterned resonators have been reported to the best of our knowledge. In addition, our metaplasmonic film shares some commonalities with the plasmonic wire medium which also uses extreme dilution\cite{pendry1996extremely,pendry1998low}, however these structures have mostly been studied at low frequencies where the metal behaves as a PEC and the effective plasma frequency becomes purely geometry dependent. Our metaplasmonic film thus operates in a markedly different regime,  where we rely on the exact permittivity of the constituent metal which strongly affects the plasmonic behavior, and where the much smaller feature sizes require strategic scaling to minimize detrimental damping effects. 

Metaplasmonics, literally `beyond plasmonics', provides a mechanism to overcome the intrinsic spectral limitations of traditional plasmonic materials, and offers a novel, quasi-2D plasmonic modality to extend plasmonic response beyond the the limited bounds of both traditional noble metals and alternative plasmonic materials. When integrated with subwavelength-scale NR geometries, these metaplasmonic films demonstrate dramatic increases in confinement factor and field-enhancement, 
opening up the previously inaccessible SWIR and mid-IR wavelength ranges to materials with true plasmonic response. These principles can be also applied to any other metals and conductive materials such as \ce{Ag} or graphene. The metaplasmonic approach using noble metals significantly outperforms other plasmonic materials in the NIR and lower mid-IR ranges, filling the technological gap existing in this range, with the potential to impact a broad range of applications including sensing, thermal manipulation, and communications. As important, this approach represents a more scalable, lower cost, and design-able off-the-shelf competitor to true 2D materials for plasmonic confinement at the extreme limits, circumventing the confinement-loss trade-off present in atomically thin metal films while providing a significantly more accessible test-bed for nonlinear and quantum plasmonics compared to graphene or other van der Waals materials.

\section{Funding Sources}
The authors acknowledge the support from the Defense Advanced Research Projects Agency (DARPA) under the Optomechanical Thermal Imaging (OpTIm) program (HR00112320022) and the National Science Foundation DMREF program (Award No. 2119302).

\bibliography{Metaplasmonics}

\end{document}